\documentclass[twocolumn,showpacs,superscriptaddress,preprintnumbers,amsmath,amssymb,epsfig,floatfix,stfloats,prb]{revtex4-1}
\usepackage{graphicx}
\usepackage{dcolumn}
\usepackage[usenames,dvipsnames]{color}
\usepackage{bm}
\usepackage{ulem}
\usepackage{natbib}
\begin{document}
\title{Substantial reduction of Stone-Wales activation barrier in fullerene}    
\author{Mukul Kabir}
\altaffiliation{Corresponding author: mukulkab@mit.edu}
\affiliation{Department of Materials Science and Engineering, Massachusetts Institute of Technology, Cambridge, Massachusetts 02139, USA}
\author{Swarnakamal Mukherjee}
\affiliation{Advanced Materials Research Unit and Department of Materials Science, S.N. Bose National Center for Basic Sciences, JD Block, Sector III, Salt Lake City, Kolkata 700 098, India}
\author{Tanusri Saha-Dasgupta}
\affiliation{Advanced Materials Research Unit and Department of Materials Science, S.N. Bose National Center for Basic Sciences, JD Block, Sector III, Salt Lake City, Kolkata 700 098, India}
\date{\today}

\begin{abstract}
Stone-Wales transformation is a key mechanism responsible for growth, transformation, and fusion in fullerene, carbon nanotube and other carbon nanostructures.  These topological defects also substantially alter the physical and chemical properties of the carbon nanostructures. However, this transformation is thermodynamically limited by very high activation energy ($\sim$ 7 eV in fullerene), which can get reduced due to the presence of hydrogen, extra carbon atom, or due to endohedral metal doping. Using first-principles density functional calculations, we show that the substitutional boron doping substantially reduces the Stone-Wales activation barrier (from $\sim$ 7 eV to 2.54 eV). This reduction is the largest in magnitude among all the mechanisms of barrier reduction reported till date. Analysis of bonding charge density and phonon frequencies suggests that the bond weakening at and around the active Stone-Wales site in B-heterofullerene is responsible for such reduction. The formation of Stone-Wales defect is, therefore, promoted in such heterofullerenes, and is expected to affect their proposed H$_2$ storage properties. Such substitutional doping can also modify the Stone-Wales activation barrier in carbon nanotube and graphene naonostructures, and catalyze isomerization, fusion, and nanowelding processes. 
\end{abstract}
\maketitle

\section{{Introduction}}
Stone-Wales (SW) defects~\cite{Stone1986501,*WalesNature1998} are important topological defects in $sp^2$-bonded carbon materials, and play crucial role in growth, isomerization, and nanoscale plasticity of carbon nanotube, fullerene, and graphitic nanostructures.~\cite{Eggen04051996, PhysRevLett.81.4656,*PhysRevB.57.R4277,PhysRevLett.88.065501} SW transformation involves an in-plane 90$^{\circ}$ bond rotation with respect to the bond center. This leads to pentagon-heptagon defects (5-7-7-5 dislocation dipole) in carbon nanotube and graphene,~\cite{PhysRevLett.81.4656,*PhysRevB.57.R4277, PhysRevB.69.121413, PhysRevLett.88.065501}  and interchange of a pair of pentagonal and hexagonal rings in the fullerene.~\cite{Stone1986501,*WalesNature1998} Such structural transformations are  believed to be the fundamental processes for the coalescence of fullerene~\cite{PhysRevLett.90.065501, *PhysRevLett.88.185501} and nanotube,~\cite{PhysRevLett.92.075504} the formation of molecular junctions for nanoelectronic devices,~\cite{MinOuyang01052001, *PhysRevLett.79.2093} and plastic deformation.~\cite{PhysRevLett.81.4656,*PhysRevB.57.R4277} Both chemically modified and unmodified SW defects induce local curvature to otherwise planer graphitic materials,~\cite{PhysRevLett.92.225502, jp710547x, Terrones19921251, PhysRevB.80.033407} which may enhance the formation of nanotube and fullerene from planer carbon nanostructures.~\cite{Nature.359} Coalescence of fullerene and nanotube has been also proposed to occur through a sequence of such structural transformation.~\cite{PhysRevLett.90.065501, *PhysRevLett.88.185501, PhysRevLett.92.075504} 
SW defects alter the electronic properties of carbon nanostructures, and thus substantially modify chemical reactivity toward adsorbates (reactivity increases compared to pristine counterpart),~\cite{Slanina200057, cr050569o, PhysRevLett.91.105502, jp0440636} and transport properties.\cite{PhysRevB.77.115453, nl802234n}  Similar SW defects are also observed in boron nitride nanotube and nanosheet,~\cite{PhysRevB.69.121413, PhysRevB.65.041406, *ct900388x} and found to have important implications in determining physical and chemical properties.~\cite{jp105454w, *ja803245d, *WonJCP2006, *jp077115a, *jp072443w} 

\begin{figure}[!t]
\begin{center}
\includegraphics[width=8cm, keepaspectratio, angle=0]{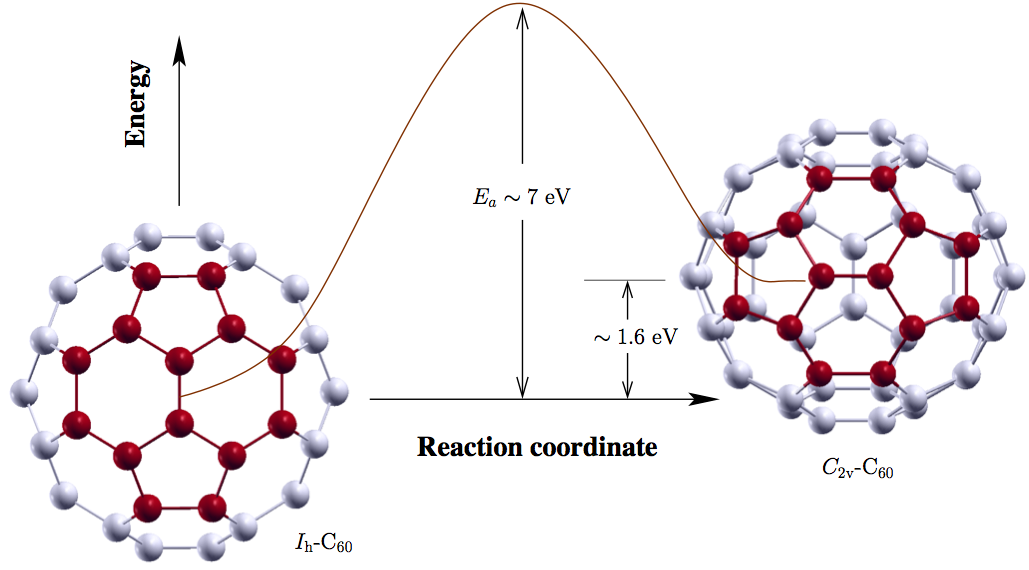}
\caption{\label{cartoon}(color online) Activation energy barrier  $E_a$ for $I_{\rm h}$-C$_{60}$  $\rightarrow$ $C_{\rm 2v}$-C$_{60}$ transition, which represents single SW transition, is very high. Activation barrier for  the reverse transition ($C_{\rm 2v}$-C$_{60}$ $\rightarrow$  $I_{\rm h}$-C$_{60}$) is smaller by $\sim$ 1.6 eV, which is the energy difference between C$_{60}$ in $I_{\rm h}$- and $C_{\rm 2v}$-symmetry. The pyracyclene region is highlighted (red).}
\end{center}
\end{figure}

It is known that 94\% of the fullerene C$_{60}$ isomers can be derived by a sequence of SW transitions.~\cite{Austin1995146} Thus, SW transformation is believed to be the mechanism for fullerene isomerization or $I_{\rm h}$-C$_{60}$ formation, which represents the global minimum on the potential energy surface, during high temperature annealing.~\cite{walsh:6691, PhysRevLett.72.669} Interchange of a pair of pentagonal and hexagonal rings in $I_{\rm h}$-C$_{60}$ through single SW transition leads to $C_{\rm 2v}$ symmetry, which does not obey the isolated-pentagon rule.~\cite{Kroto.Nature} This is the first isomer and separated by $\sim$ 1.6 eV energy from the global $I_{\rm h}$-minimum.~\cite{j100020a035, PhysRevB.73.113406, yi:8634} Therefore, $C_{\rm 2v}$-C$_{60}$ represents the first step toward fullerene isomerization, or the last step of annealing process before reaching the icosahedral global minimum.~\cite{walsh:6691} Although the pristine $C_{\rm 2v}$ isomer is not experimentally observed earlier, the chlorinated $C_{\rm 2v}$ isomer has recently been stabilized experimentally via Kr\"{a}tschmer-Huffman synthesis,~\cite{NatMat.Tan} and in the gas phase by subsequent dechlorination.~\cite{jp902325w} The SW transition in fullerene is, however, thermodynamically limited due to very large activation barrier ($\sim$ 7 eV, Fig.~\ref{cartoon}).~\cite{Dresselhaus.Book, PhysRevB.73.113406, ja0288744} The barrier height is found to get reduced in the presence of an extra carbon or hydrogen atom.~\cite{Eggen04051996, Heggie.Book} The extra carbon atom acts as autocatalyst and reduces activation energy by $\sim$ 2 eV, when placed preferentially in the regions of paired pentagons. In contrast, the endohedral metal doping (K, Ca, and La) has found to be relatively less effective in barrier height reduction.~\cite{PhysRevB.73.113406} 

Here we report the effect of substitutional doping on SW transition in C$_{60}$ through first-principles density functional calculations. We find that the activation barrier is reduced substantially by substitutional boron doping at the active SW sites. The magnitude of reduction (2.12 eV and 4.34 eV for single and double B-doping, respectively) is much larger compared to the cases with extra carbon or hydrogen or endohedral metal doping.~\cite{Eggen04051996, Heggie.Book, PhysRevB.73.113406} Boron doped heterofullerenes (C$_{60-x}$B$_{x}$, $x \leqslant 6$) have been successfully synthesized,~\cite{Muhr1996399, *j100166a010} and are believed to be a promising candidate for hydrogen storage.~\cite{PhysRevLett.94.155504, PhysRevLett.96.016102}  It has been shown that H$_2$ adsorption energy increases substantially for B-heterofellerenes, where the B-centers act as adsorption centers.  Our study shows that formation of SW defects are promoted in such B-doped heterofullerenes, which may alter the reversible storage properties as SW defects are known to be more reactive towards adsorbates.  Presence of such substitutional dopants is also expected to reduce the activation barrier and promote SW transition in other carbon nanostructures, and thus catalyze isomerization, fusion, and nanowelding processes.

\section{Methodology}
Calculations were carried out using density functional theory implemented in the Vienna Ab-initio Simulation Package (VASP) code~\cite{PhysRevB.47.558, *PhysRevB.54.11169} with the Perdew-Burke-Ernzerhof exchange-correlation functional~\cite{PhysRevLett.77.3865} and projector augmented wave pseudopotential~\cite{PhysRevB.50.17953}. The kinetic energy cutoff was chosen to be 800 eV. Symmetry unrestricted geometry optimizations were terminated when the force on each atom was less than 0.005 eV/\AA, and the reciprocal space integrations were carried out at the $\Gamma$ point. We determined the minimum-energy path for SW transition and the corresponding migration energy barrier $E_a$ using climbing-image nudged elastic band (NEB) method.~\cite{henkelman:9901, *henkelman:9978} In NEB, a set of intermediate states (images) are distributed along the reaction path connecting optimized initial ($I_{\rm h}$-C$_{60}$) and final ($C_{\rm 2v}$-C$_{60}$) states. To ensure the continuity of the reaction path, the images are coupled with elastic forces and each intermediate state is fully relaxed in the hyperspace perpendicular to the reaction coordinate. 

\section{Results and discussions}
Determination of the activation energy necessary for SW transition ($I_{\rm h}$-C$_{60}$ $\rightleftharpoons$ $C_{\rm 2v}$-C$_{60}$) via climbing-image NEB method requires structural information of  the initial and final structures, within the concerned level of theory.  Therefore, before we discuss the effect of B-doping on the activation energy, we begin our discussion with the structure and electronic properties of C$_{60}$, C$_{59}$B, and C$_{58}$B$_2$ cages.

\begin{figure}[!bp]
\begin{center}
\includegraphics[width=8cm, keepaspectratio, angle=0]{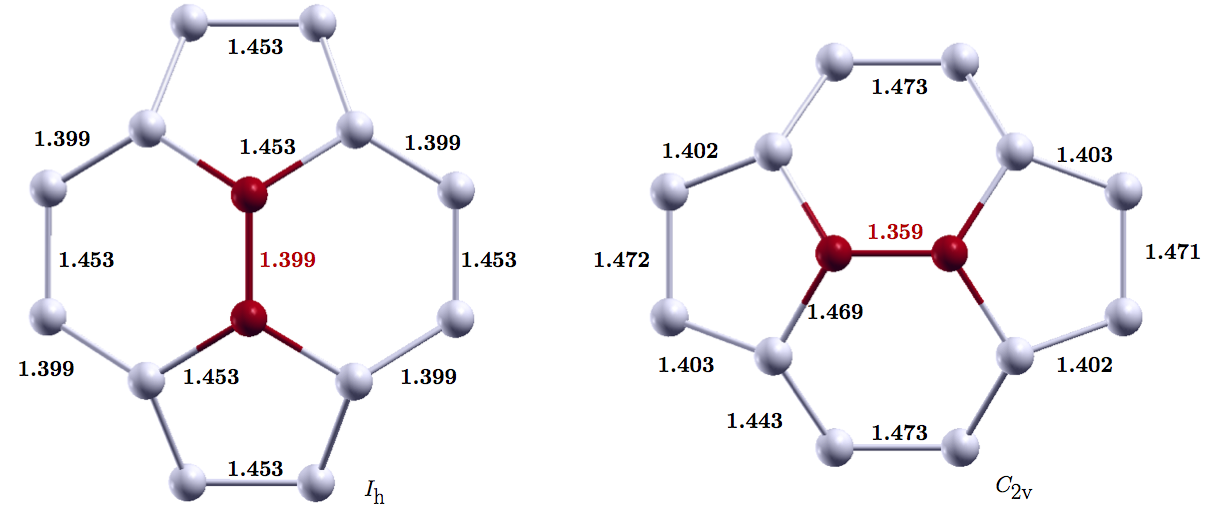}
\caption{\label{pyra-bond}(color online) Local pyracyclene region for the optimized C$_{60}$ in $I_{\rm h}$ and $C_{\rm 2v}$ symmetries. The rotating dimer corresponding to the SW transition is highlighted, and the bond lengths are shown in \r A. The rotating bond length decreases substantially in the $C_{\rm 2v}$ structure, and induces a significant local distortion. Similar trend is also observed in B-heterofullerenes.~\cite{supple}}
\end{center}
\end{figure}

\subsection{C$_{60}$, C$_{59}$B, and C$_{58}$B$_2$: Structure and Electronic properties}
\begin{figure*}[!t]
\begin{center}
\includegraphics[width=17cm, keepaspectratio, angle=0]{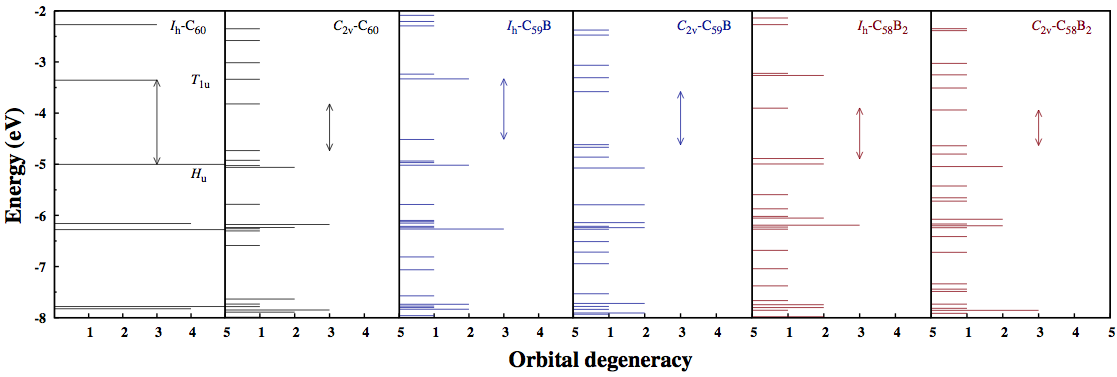}
\caption{\label{levels}(color online) Orbital degeneracy for pristine and substitutional B-doped fullerenes (C$_{59}$B and C$_{58}$B$_2$) for both $I_{\rm h}$ and $C_{\rm 2v}$ symmetries. The HOMO (LUMO) of the icosahedral C$_{60}$, $H_u$ ($T_{1u}$) is five (three) fold degenerate. Bottom (top) of the vertical arrow represents HOMO (LUMO) for the corresponding structure. Introduction of defects, both in terms of topological defect or substitutional B, breaks the orbital degeneracy, and thus reduces the energy gap $E_g$ (see Table~\ref{table-energy}).}
\end{center}
\end{figure*}

There are two types of bonds in $I_{\rm h }$-C$_{60}$: [6,6]-bonds (1.399 \r A) at the junctions of two hexagons are smaller than [5,6]-bonds (1.453 \r A) at the junctions of a pentagon and a hexagon. The optimized geometries in both $I_{\rm h }$ and $C_{\rm 2v}$ symmetries are shown in Fig.~\ref{cartoon}, and 14-atom pyracyclene region containing the rotating C$_2$ dimer is shown in Fig.~\ref{pyra-bond}. Calculated [5,6] and [6,6] bond lengths (Fig.~\ref{pyra-bond}) are in excellent agreement with previous theoretical calculations~\cite{ja0288744, Haser1991497} and experimental measurements~\cite{David.Nature.353, KENNETHHEDBERG10181991, SHENGZHONGLIU10181991, ja00008a068} for the $I_{\rm h}$ structure. The rotating bond in $C_{\rm 2v}$-C$_{60}$ is much smaller, 1.359 \r A, which is also in agreement with previous theoretical calculations.~\cite{ja0288744, Haser1991497, yi:8634} However, no experimental information is available on pristine $C_{\rm 2v}$-C$_{60}$, chlorinated $C_{\rm 2v}$ structure (C$_{60}$Cl$_8$) shows a similar bond shortening of the rotating dimer (1.37 \r A).~\cite{NatMat.Tan} As shown in Fig.~\ref{pyra-bond}, such bond shortening introduces local strain, and thus a distribution of bond lengths is observed in the pyracyclene region.   

When one of the C atoms in the rotating dimer (see supplementary document~\cite{supple}) is replaced by B, three bond lengths involving C and B  are increased (two [5,6]-C$-$B=1.549 \r A, and [6,6]-C$-$B=1.524 \r A) for $I_{\rm h}$-C$_{59}$B. This is responsible for the bond deformation in pyracyclene region (the [5,6]-C$-$C bond distribution ranges from 1.442 \r A to 1.482 \r A). The trend of the structural changes in moving from $I_{\rm h}$-C$_{59}$B to $C_{\rm 2v}$-C$_{59}$B, remains same as is observed for pristine C$_{60}$. Although, the structural properties of C$_{58}$B$_{2}$ show similar qualitative behavior, the pyracyclene region is comparatively more deformed due to the presence of long B$-$B (1.663 \r A) bond, and four longer C$-$B (1.551 \r A) bonds. Though, strictly speaking, both the icosahedral and $C_{\rm 2v}$ symmetries are broken due to B-doping, for simplicity we shall retain the same nomenclature in the following.

\begin{table}[!b]
\caption{\label{table-energy} Energy difference between the $I_{\rm h}$ and $C_{\rm 2v}$ symmetries, $\Delta E = E(I_{\rm h}) - E(C_{\rm 2v})$; and the energy gap $E_g$ between the highest occupied molecular orbital (HOMO) and lowest unoccupied molecular orbital (LUMO) for both $I_{\rm h}$- and $C_{\rm 2v}$-symmetry. All the energies are in eV.}
\begin{tabular}{lcccccc}
\hline
            & \ \ \ & $\Delta E$    & \ \ \ & $E_{g} (I_{\rm h})$  & \ \ \ &  $E_{g} (C_{\rm 2v})$ \\
             \hline
             \hline
Pristine-C$_{60}$ &  &  1.57   & & 1.63  & & 0.91 \\             
C$_{59}$B         &  &  1.67   & & 1.18  & & 1.04  \\
C$_{58}$B$_2$     &  &  1.75   & & 0.98  & & 0.70   \\
\hline
\end{tabular}
\end{table}

The $C_{\rm 2v}$ structure is the first isomer, and in agreement with previous theoretical reports,~\cite{j100020a035, PhysRevB.73.113406, yi:8634,ja0288744} we find that this structure is 1.57 eV higher in energy compared to the global $I_{\rm h}$-minimum. Calculated energy gap between the highest occupied molecular level (HOMO) and the lowest unoccupied molecular level (LUMO) is 1.63 eV for $I_{\rm h}$-symmetry, which agrees well with the spectroscopic measurements.~\cite{Yang1987233, PhysRevLett.81.5378} Due to the icosahedral symmetry, the HOMO ($H_{u}$) and LUMO ($T_{1u}$) are five and three fold degenerate, respectively, in $I_{\rm h}$-C$_{60}$. These degeneracies are lifted (Fig.~\ref{levels}) due to the presence of topological defect introduced by SW transition or due to the substitutional B-doping. Substitutional B introduces a local bond deformation in the structure, and also introduces holes into the system. 
Both causes reduction of the HOMO-LUMO gap, which are shown in Table~\ref{table-energy}. For pristine $C_{\rm 2v}$-C$_{60}$, the degeneracy of both HOMO and LUMO is completely broken (Fig.~\ref{levels}), which in turn reduces $E_g$ by $\sim$ 0.7 eV.

Single substitutional B introduces a significant bond deformation,~\cite{supple} and thus the inherent symmetry of C$_{59}$B deviates from perfect $I_{\rm h}$-symmetry, which removes the five-fold degeneracy of the HOMO level (Fig.~\ref{levels}). Substitutional B introduces hole (one/B-atom) into the system, and thus the HOMO in $I_{\rm h}$-C$_{59}$B is singly occupied, which is pushed 0.47 eV above compared to the pristine $I_{\rm h}$-C$_{60}$, which reduces the gap. In addition, when a SW topological defect is introduced the three-fold degeneracy of LUMO is completely lifted (Fig.~\ref{levels}), and pulled down by $\sim$ 0.2 eV. This reduces the gap even more for the $C_{\rm 2v}$-C$_{59}$B structure. Similarly, when two C atoms are replaced (the system has two holes) the degeneracy in HOMO is broken. One of these levels is completely unoccupied and serves as LUMO in $I_{\rm h}$-C$_{58}$B$_2$.  This level lies 0.98 eV above the occupied HOMO. The orbital degeneracy of the HOMO is further broken and pushed above to further reduce the gap in $C_{\rm 2v}$-C$_{58}$B$_2$ structure. 

\subsection{Activation barrier: Effect of substitutional B-doping}
\begin{figure}[!t]
\begin{center}
\includegraphics[width=8.5cm, keepaspectratio, angle=0]{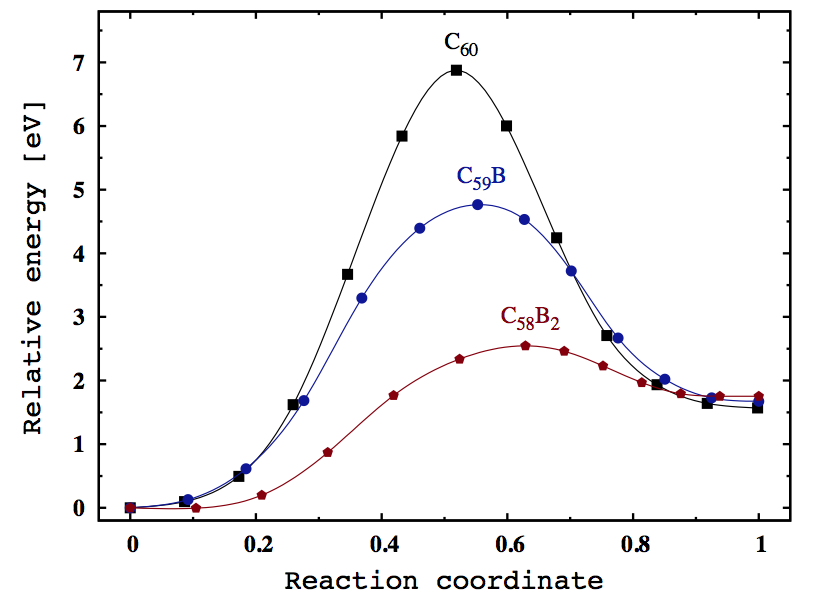}
\caption{\label{neb}(color online) The activation energy barrier to the SW transition ($I_{\rm h} \rightarrow C_{\rm 2v}$) reduces substantially for the B-heterofullerenes, when B is doped in the active SW site (rotating dimer). The reverse barrier is smaller by an amount $\Delta E$, reported in Table~\ref{table-energy}, which are similar in magnitude for all compositions. 
}
\end{center}
\end{figure}

Activation barriers for the SW transition in carbon nanotube, graphitic nanostructure, and fullerene are generally very high.~\cite{PhysRevLett.61.2693,dumitrica:2775, yi:8634} 
Our calculated energy barrier for the pristine C$_{60}$ is found to be 6.88 eV (Fig.~\ref{neb}), which is in agreement with previous reports obtained within a range of theories.~\cite{ja0288744, PhysRevB.73.113406, yi:8634, Eggen04051996} Compared to carbon nanotube or graphitic nanostructures, the activation barrier  in fullerene is considerably less due to the strain present in fullerene structure.~\cite{yi:8634} We find that the rotating C$_2$ unit in the transition state (TS) pops-out (by $\sim$ 0.21 \r A) from the C$_{60}$ surface. Similar TS was predicted previously.~\cite{Murry.Nature, yi:8634, ja0288744} The bond length of the rotating C$_2$ unit shrinks to 1.248 \r A in the TS from the corresponding bond length in $I_{\rm h}$ (1.399 \r A) and $C_{\rm 2v}$ (1.359 \r A) symmetries. Therefore, this C$_2$ bond can be characterized as a stretched $sp^3$-bond.~\cite{ja0288744}

We find that the large SW activation barrier is reduced substantially by B-doping (Fig.~\ref{neb}) at the active SW site. 
The barrier is reduced by 2.12 eV, when a single C-atom is replaced in the rotating dimer by B for the C$_{59}$B heterofullerene. The reduction is comparable in magnitude with the cases where an extra carbon was strategically placed in the regions of paired pentagons~\cite{Eggen04051996} or catalyzed by hydrogen.~\cite{Heggie.Book} In contrast, the endohedral metal doping has relatively smaller effect. The barrier was found to reduce by only 0.80 eV for La-doping.~\cite{PhysRevB.73.113406} In the present case, the barrier is further reduced to 2.54 eV, when both the C-atoms in the rotating dimer are replaced with B-atoms for C$_{58}$B$_2$ heterofullerene. Compared to pristine C$_{60}$, a total reduction of 4.34 eV is obtained which is the maximum reported till date.  
We find that for pristine C$_{60}$ as well as B doped C$_{60}$, the structure corresponding to the saddle point  has only one vibrational mode with imaginary frequency. This confirms that the obtained saddle points are indeed true first-order transition states. We find that the frequency of this imaginary mode decreases with the introduction of single B, which further decreases with the introduction of second B (962 cm$^{-1}$, 377 cm$^{-1}$, and 167 cm$^{-1}$ for C$_{60}$,  C$_{59}$B, and  C$_{58}$B$_2$, respectively). Thus the variation of energy along the reaction path becomes slower, i.e, the NEB curve gets flatter, with the introduction of boron (Fig.~\ref{neb}).

Origin of the reduction in SW activation barrier can be explained in terms of changes in bond strength, charge density and phonon spectra, which point toward bond weakening at and around the rotating dimer. First, we calculate C$-$C, C$-$B, and B$-$B bond strengths in pristine C$_{60}$, C$_{59}$B, and C$_{58}$B$_2$, respectively, which can be approximately calculated through following expressions,  
\begin{eqnarray}
E_{\rm C-C} &=& E^{\rm tot}_{{\rm C}_{60}} / 90  
\nonumber \\
E_{\rm C-B} &=& \left[ E^{\rm tot}_{{\rm C}_{59}{\rm B}} - 87 \times  E_{\rm C-C} \right] / 3
\nonumber \\
E_{\rm B-B} &=& \left[ E^{\rm tot}_{{\rm C}_{58}{\rm B}_2} - 85 \times  E_{\rm \tiny C-C} - 4 \times  E_{\rm C-B} \right], 
\end{eqnarray}
where $E^{\rm tot}$ represents the total binding energy of the respective structures. Calculated bond strength decreases with B-doping: 5.87, 4.74 and 3.75 eV, respectively, for $E_{\rm C-C}$, $E_{\rm C-B}$ and $E_{\rm B-B}$. This hints toward the presence of weaker C$-$B bonds around the rotating dimer, causing the SW bond rotation to be easier for the B-doped heterofullerenes. This qualitatively explains the reduction in activation barrier. We further note that the amount of reduction is related to the reduction in bond strength as $\delta E_a = n [E_{\rm C-C} - E_{\rm C-B}]$, where $n$ is the number of C$-$B bonds present in C$_{59}$B (two), and C$_{58}$B$_2$ (four) heterofullerenes.

\begin{figure}[!t]
\begin{center}
\includegraphics[width=8.5cm, keepaspectratio, angle=0]{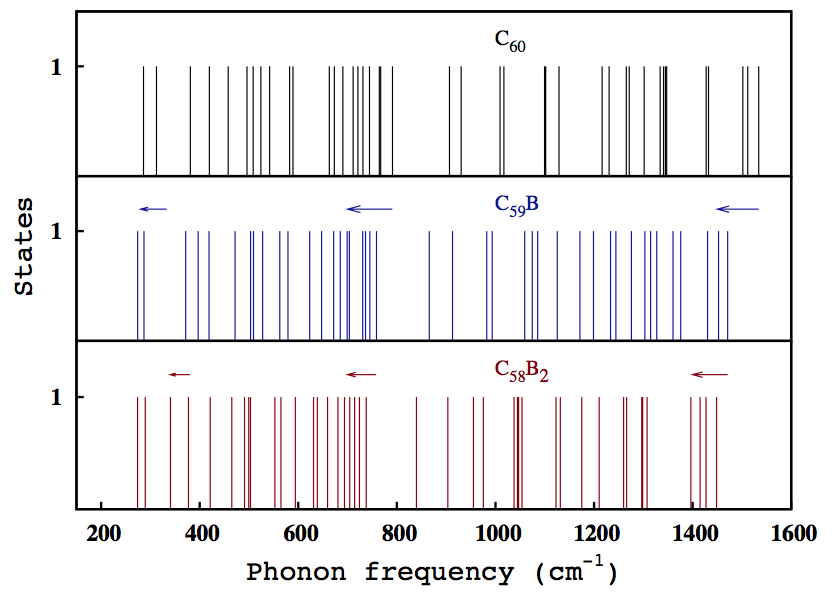}
\caption{\label{phonon}(color online) Due to the B-doping phonons shift toward lower frequencies, which indicate bond weakening. Phonons for the 14-atom pyracyclene around the rotating dimer is shown for $I_{\rm h}$-symmetry. Arrows are guide to eye.}
\end{center}
\end{figure}

\begin{figure*}[!t]
\begin{center}
\includegraphics[width=14cm, keepaspectratio, angle=0]{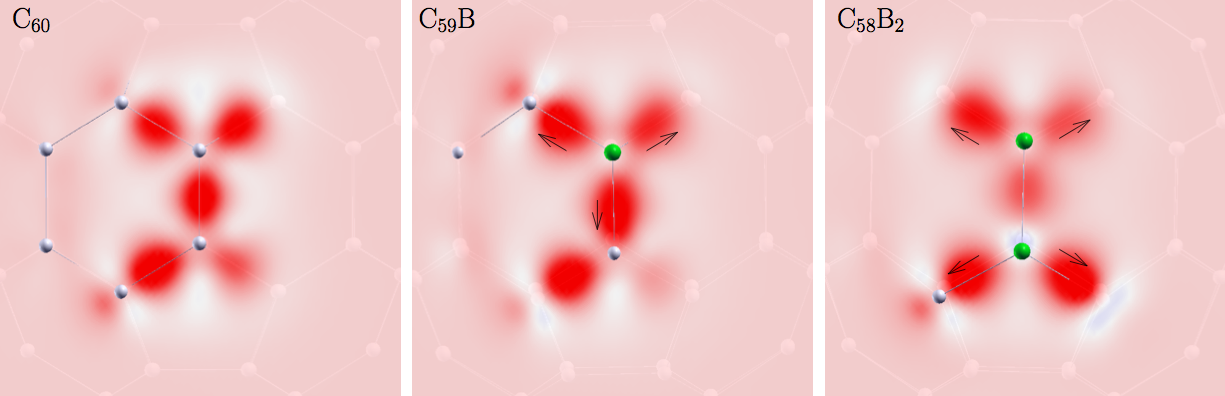}
\caption{\label{cdensity}(color online) Bonding charge densities  $\Delta \rho$ for pristine C$_{60}$, and single and double B-doped heterofullerenes. The B atoms are shown in green.  The $\Delta \rho$  for a particular cage is calculated as, $\Delta \rho = \rho^{\rm tot}(\mathbf r) - \rho^{\rm C_{58}}(\mathbf r) - \rho^{\rm X}(\mathbf r) - \rho^{\rm Y}(\mathbf r)$, where X$-$Y represents the rotating  C$-$C,  C$-$B and B$-$B dimers for C$_{60}$, C$_{59}$B, and C$_{59}$B$_2$ structures, respectively. For pristine C$_{60}$, the charge accumulation centers (red) appear at the middle of the bond leading to C$-$C covalent bond, which shifts significantly toward the C-atoms (shown with the arrows) due to B-doping leading to ionic character of the C$-$B bonds.}
\end{center}
\end{figure*}

Next we turn our attention to the phonon frequencies calculated at the $\Gamma$ point. It is reasonable to assume that the atoms around the rotating bond 
are mostly affected due to B-doping and SW rotation. Thus, we considered the 14-atom pyracyclene ring around the rotating bond to calculate the normal vibrational modes. This significantly reduces the size of Hessian matrix (from 180 $\times$ 180 for the full cluster to 42$\times$42) to calculate phonon frequencies. Fig.~\ref{phonon} shows that due to B-doping, phonon frequencies shift gradually to the lower frequencies, indicating the bonds get softer 
as the phonon frequency ($\nu$) is related to  bond stiffness ($k$) as, $\nu \propto \sqrt k$. The observed softening in phonon frequencies arises 
due to two factors. Firstly, the C$-$B and B$-$B bonds are longer and also less stronger compared to the C$-$C bonds, as pointed out earlier. Additionally, the C$-$C bonds around the rotating unit also get elongated due to the substantial strain introduced by B-doping, and thus get weakened. The rotation of bond is easier in such softer environment, which eventually reduces the SW activation barrier. 

Bonding charge density analysis  (Fig.~\ref{cdensity}) also qualitatively points towards softening of the bonds that are involved in SW process. Figure~\ref{cdensity} shows the plots of bonding charge densities for pristine C$_{60}$, singly and doubly B-doped C$_{60}$. For  pristine C$_{60}$, the C$-$C 
bonding is completely covalent i.e., the bonding charge accumulation is at the centers of the bond.  This picture deviates for B-doped cases, where the charge accumulation centers move toward the C-atom for C$-$B bonds, imparting ionic character to the C$-$B bonds. Thus, the C$-$C rotation is increasingly easier for C$_{59}$B and C$_{58}$B$_2$ heterofullerenes. The bonding character can be also quantitatively understood from Bader charge analysis,~\cite{Bader, *Henkelman2006354} which reveals that each B-atom looses $\sim 1 |e|$ charge to the neighboring C-atoms. This indicate deviation from perfect covalent character.

\subsection{Reaction rate and characteristic time-scale}

\begin{figure}[!hb]
\begin{center}
\includegraphics[width=8cm, keepaspectratio, angle=0]{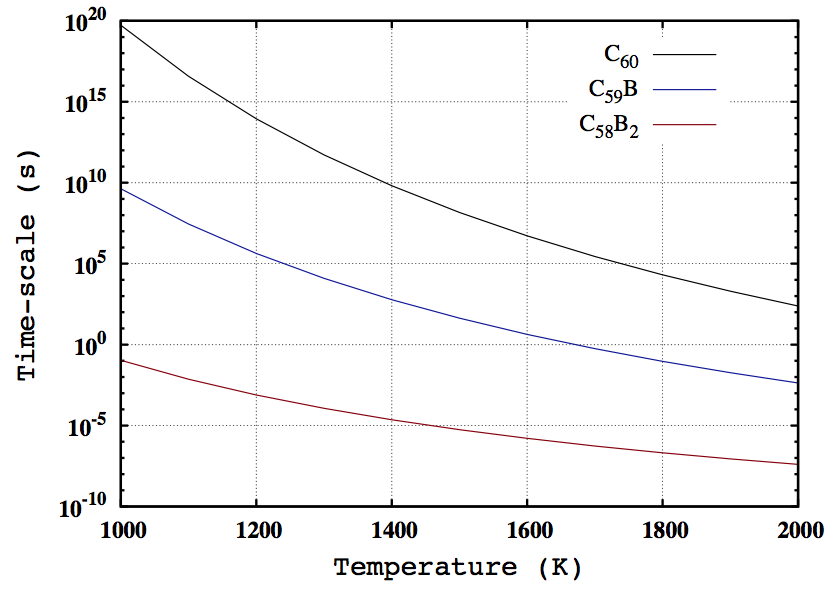}
\caption{\label{rate}(color online)  The $I_{\rm h} \rightarrow C_{\rm 2v}$ transition (reaction) rate, and thus the characteristic time scale for the corresponding transition has Arrhenius dependence on temperature and activation energy. Subsequent B-doping in the rotating dimer reduces the characteristic time-scale enormously via decrease in the activation barrier.}
\end{center}
\end{figure}

At temperatures well below the melting point, the harmonic approximation to transition state theory (HTST) can be applied to study the reaction rate or characteristic time-scale associated with the reaction.   The reaction rate can be expressed in terms of energy and normal mode frequencies at the saddle point and initial state,~\cite{Vineyard1957121} 
\begin{equation}
k^{\rm HTST} = \frac{\displaystyle\prod_i^{3N}\nu_i^{\rm I}}{\displaystyle\prod_i^{3N-1}\nu_i^{\rm TS}} e^{-(E_a/k_BT)},
\end{equation}
where $E_a$ = ($E^{\rm TS} - E^{\rm I}$) is the activation energy with $E^{\rm I}$ ($E^{\rm TS}$) being the energy corresponding to initial (transition) state, and the $\nu_i$ are the corresponding normal mode frequencies. Within the harmonic approximation, the the entropic effect to the reaction rate, is 
included in the prefactor involving phonon frequencies calculated at zero temperature.  The first-order transition state is characterized by one imaginary phonon, which is excluded  for the transition state. To calculate the prefactor we have only considered the normal modes corresponding to 14-atom pyracyclene ring around the rotating dimer. The characteristic time-scale associated with a given reaction can be calculated from the knowledge of reaction rate $k^{\rm HTST}$ as $\tau = 1/k^{\rm HTST}$. Figure~\ref{rate} shows the characteristic time required for single $I_{\rm h} \rightarrow C_{\rm 2v}$ SW transition plotted as a function of temperature. The characteristic time scale, measured in second, differs by $10^{10} - 10^6$ between C$_{60}$, C$_{60}$B and C$_{60}$B$_2$ in the temperature range 1000 - 2000 K. For example, at 1700 K, single SW bond rotation event takes place in $\sim$ 10 days for pristine C$_{60}$. In contrast, at the same temperature, the process is enormously accelerated  and requires only time scale of about second and micro-second for C$_{59}$B and C$_{58}$B$_2$, respectively. It should be pointed out that the reverse transition  ($C_{\rm 2v} \rightarrow I_{\rm h}$, which is not shown in Fig.~\ref{rate}) is much faster than the forward transition ($I_{\rm h} \rightarrow C_{\rm 2v}$, which is shown in Fig.~\ref{rate}). This is caused by the presence of exponential factor, $\exp(-\Delta E/k_BT)$, which is five orders of 
magnitude smaller for $C_{\rm 2v} \rightarrow I_{\rm h}$, compared to $I_{\rm h} \rightarrow C_{\rm 2v}$ at temperature 1700 K.

\section{Conclusions}
The Stone-Wales activation barrier in pristine fullerene is very large. Earlier it has been found that this high barrier can get reduced by 
$\sim$ 35\% by the presence of an extra carbon or hydrogen.~\cite{Eggen04051996, Heggie.Book} Our present study shows that presence of substitutional
boron at the active Stone-Wales site can become even more effective in reduction of the Stone-Wales activation barrier. It can get reduced substantially by $\sim$ 30\% and 60\%, 
due to single and double B-doing, respectively. Calculated charge density and phonon spectrum indicate that the C$-$B bonds are softer compared to the C$-$C bonds, 
and the presence of such weaker bonds around the rotating dimer is responsible for the observed reduction in activation barrier.  From 
thermodynamic point of view, such reduction in the activation barrier $\Delta E_a$ enormously reduces the time-scale associated with the Stone-Wales process. The reduction is of the order of $\sim \exp(\Delta E_a/k_BT) \sim 10^{6}$  $(10^{12})$ fold at temperature 1700 K for single (double) B-doping.  Stone-Wales defects are known to alter the 
chemical reactivity toward adsorbates,~\cite{Slanina200057, cr050569o, PhysRevLett.91.105502, jp0440636} thus are expected to effect the proposed reversible 
H$_2$ storage properties of B-heterofullerenes.~\cite{PhysRevLett.94.155504, PhysRevLett.96.016102}  Though, here we have shown the reduction of Stone-Wales 
activation barrier for fullerene, similar reduction is expected for carbon nanotube and graphene nanostructures due to B-doping, which will catalyze fusion, 
and welding process necessary for nanoelectronic device applications.

\begin{acknowledgements}
M. K. acknowledges congenial hospitality of S. N. Bose National Centre for Basic Sciences, Kolkata. S. M. thanks Council of Scientific and Industrial Research, India for financial support.
 \end{acknowledgements}


%

\end{document}